\documentstyle[12pt,epsfig]{article}
\makeatletter

\newcommand{\PRL}[1]{{\it Phys.\ Rev.\ Lett.\ }{\bf #1}}
\newcommand{\PRD}[1]{{\it Phys.\ Rev.\ }{\bf D#1}}
\newcommand{\PLB}[1]{{\it Phys.\ Lett.\ }{\bf B#1}}
\newcommand{\ZPC}[1]{{\it Z.\ Phys.\ }{\bf C#1}}
\newcommand{\hep}[1]{[{\it hep-ph/#1}]}

\newcommand{\mrm}[1]{\mathrm{#1}}
\renewcommand{\c}{\mrm{c}}
\renewcommand{\b}{\mrm{b}}
\renewcommand{\d}{\mrm{d}}
\newcommand{\e}{\mrm{e}}

\newcommand{\p}{\mrm{p}}

\newcommand{\ccbar}{\c\overline{\mrm{c}}}

\newcommand{\bbbar}{\b\overline{\mrm{b}}}
\newcommand{\QQbar}{Q\overline{Q}}

\newcommand{\pbar}{\overline{\mrm{p}}}

\newcommand{\mc}{m_{\c}}
\newcommand{\mb}{m_{\b}}

\newcommand{\BR}{\mrm{Br}}

\newcommand{\lessim}{\raisebox{-0.8mm}%
{\hspace{1mm}$\stackrel{<}{\sim}$\hspace{1mm}}}

{\end{list}}
\newcounter{enumct}

\newlength{\abstwidth}
\begin{document}
 
\sloppy

\pagestyle{empty}

\begin{flushright}
CERN--TH/96--157  \\
LBL--39012 \\
June 1996
\end{flushright}
 
\vspace{\fill}
 
\begin{center}
{\LARGE\bf Systematics of quarkonium production}\\[10mm]
{\Large Gerhard A. Schuler$^a$} \\[3mm]
{\it Theory Division, CERN,
CH-1211 Geneva 23, Switzerland}\\[1mm]
{ E-mail: schulerg@afsmail.cern.ch}\\[2ex]
{\large and} \\[2ex]
{\Large Ramona Vogt$^b$} \\[3mm]
{\it Nuclear Science Division, Lawrence Berkeley National Laboratory,
Berkeley, CA~94720 USA\\[1mm]
and\\[1mm]
Physics Department, University of California at Davis,
Davis, CA~95616 USA}\\[1mm]
{ E-mail: vogt@nsdssd.lbl.gov}
\end{center}
 
\vspace{\fill}
 
\begin{center}
{\bf Abstract}\\[2ex]
\begin{minipage}{\abstwidth}
Quarkonium production in high-energy reactions is found to
exhibit a behaviour more universal than that expected from velocity scaling.
Total rates of quarkonia produced in hadronic interactions as
well as Feynman-$x$ and transverse momentum distributions
can be described over the full range of accessible 
energies ($15\, \lessim \sqrt{s} \lessim 1800\,$GeV) 
by two-stage processes.  The quarkonium production cross section factors 
into a process-dependent short-distance 
part and a single long-distance matrix element. 
The first part describing 
the production of a free quark--antiquark pair is the 
perturbatively calculated subthreshold cross section.
The non-perturbative factor turns out to be universal, giving the model
great predictive power.  Furthermore we estimate 
the fraction of the heavy-quark cross section leading to quarkonium
for both the charm and bottom
systems. Finally, we comment on quarkonium photoproduction.
\end{minipage}
\end{center}

\vspace{\fill}
\noindent
\rule{60mm}{0.4mm}

\vspace{1mm} \noindent
${}^a$ Heisenberg Fellow.\\[1mm]
\noindent
${}^b$ This work was supported in part by the Director,
Office of Energy Research, Division of Nuclear Physics
of the Office of High Energy and Nuclear Physics of the U. S.
Department of Energy under Contract Number DE-AC03-76SF0098.\\[10mm]
%
\noindent
CERN--TH/96--157

\clearpage
\pagestyle{plain}
\setcounter{page}{1} 
Recently much effort has been devoted to explain quarkonium production in 
a new factorization approach \cite{BFY96}. Any quarkonium cross section
is given by the sum of infinitely many terms, each of which factors 
into the product of two terms. The first one, calculable as a series
in $\alpha_s(\mu)$ where $\mu$ is of the order of the relevant hard scale, 
is the cross section to produce a free quark--antiquark pair ($Q\bar{Q}$) 
in a particular angular momentum and colour state ${}^{2S+1}L_J^{(c)}$
(in the spectroscopic notation and $c=1$ ($c=8$) denotes a colour-singlet
(colour-octet) state).
The second factor determines the probability that such a $Q\bar{Q}$ pair 
binds to form a quarkonium $H(nJ^{PC})$ of given total spin $J$, 
parity $P$, and charge conjugation $C$. The factorization approach 
becomes meaningful through the velocity-scaling rules, which determine 
the relative importance of the various long-distance matrix elements (ME). 
At any order in $v$, the velocity of the heavy quark within the bound state, 
the quarkonium cross section is hence given by a finite number of 
contributions \cite{BBL95}. 

This results in an expansion of the quarkonium cross section in both 
$\alpha_s(\mu)$ and $v$. Although this 
``velocity-scaling model'' (VSM) suggests an explanation 
of quarkonium production at the Tevatron \cite{BFY96}, its weak point 
is the fact that currently the non-perturbative ME cannot be calculated
in QCD\footnote{Attempts to calculate decay MEs on the lattice have just 
started \cite{BSK96}.}. A crucial test of the approach is therefore the 
determination of the MEs from as many different high-energy 
reactions as possible. 
This endeavour is, however, rendered more difficult by the fact 
that, in general, different combinations of MEs arise. 
Nonetheless, preliminary attempts indicate that the velocity-scaling is 
not perfect: $J/\psi$ production at the Tevatron requires considerably 
larger $c=8$ MEs \cite{CL96} than $J/\psi$ hadroproduction \cite{BR96}
or the $z$-distribution in $J/\psi$ photoproduction \cite{CK96}. 
Also the hadroproduction ratio $\chi_{\c 1} / \chi_{\c 2}$
is too low compared to data 
since $\chi_{\c 1}$ production is clearly disfavoured 
by either a power of $\alpha_s(m_c)$ or a factor $(v^2)^2$ 
compared to $\chi_{\c 2}$. Last but not least, the $J/\psi$ (non-) 
polarization is difficult to account for in the VSM \cite{BR96,Vant}.

Some time ago we \cite{Our} have shown that existing quarkonium 
production data at fixed-target energies are, in fact, compatible with the 
assumption that the non-perturbative transition of the $Q\bar{Q}$ pair to
quarkonium is more universal than expected from the velocity-scaling rules. 
Indeed, low-energy data are well reproduced if the (infinite) sum 
of short-distance coefficients times long-distance MEs is truncated to a 
single term
\begin{equation}
  \sigma\left[ H\left(nJ^{PC}\right) \right]
  = F[ nJ^{PC}]
  \,
  \tilde{\sigma}\left[ Q\bar{Q} \right]
\ .
\label{master}
\end{equation}
In this letter we shall demonstrate that the colour-evaporation model 
(CEM) of eq.~(\ref{master}) also accounts for quarkonium production at the 
Tevatron and comment upon its application to photoproducton of quarkonia.
We emphasize that in contrast
to the VSM, the number of non-perturbative parameters is minimal, hence the
model possesses great predictive power.

\begin{figure}
\begin{center}
\begin{tabular}{cc}
\mbox{\epsfig{figure=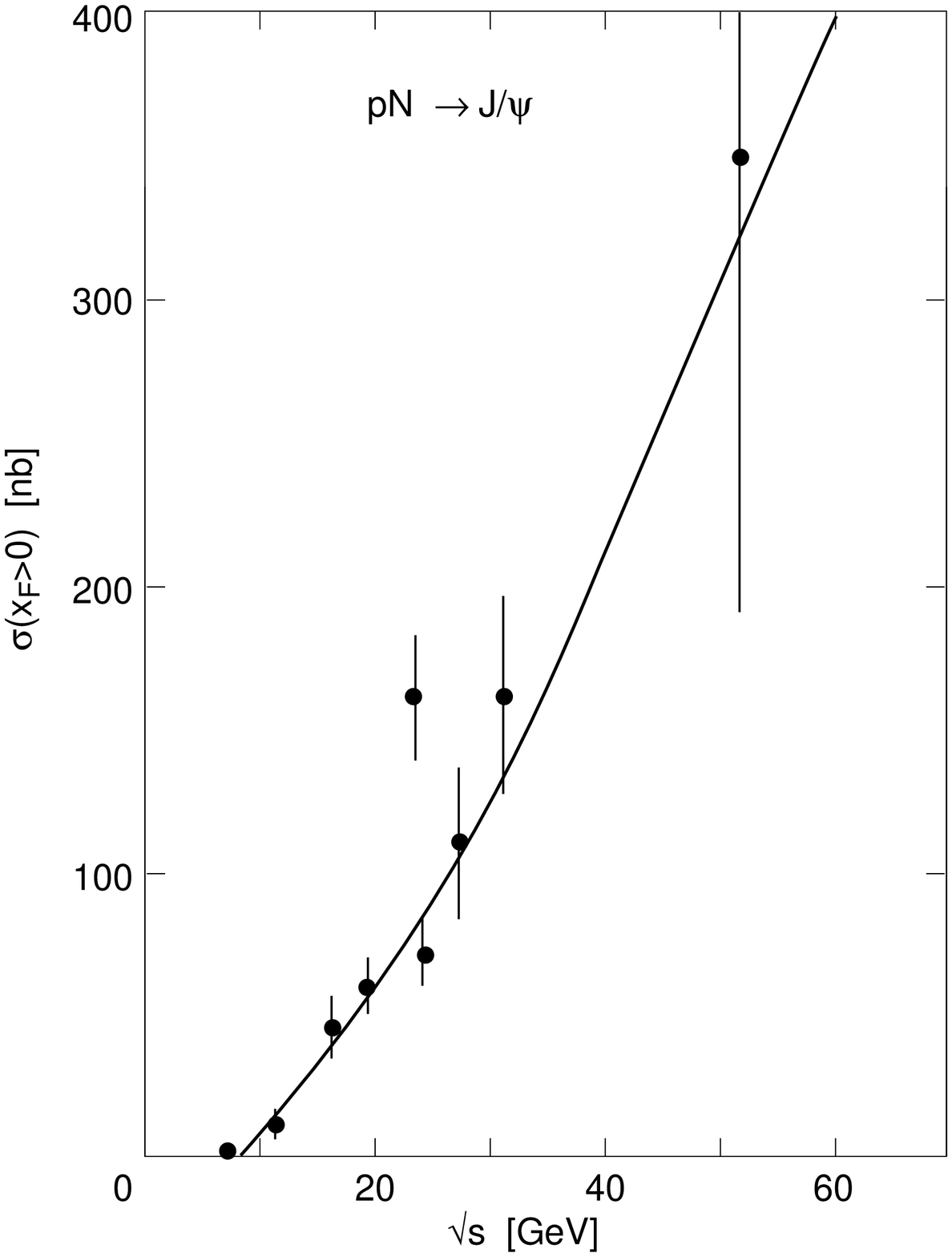,height=0.60\textwidth}}
\mbox{\epsfig{figure=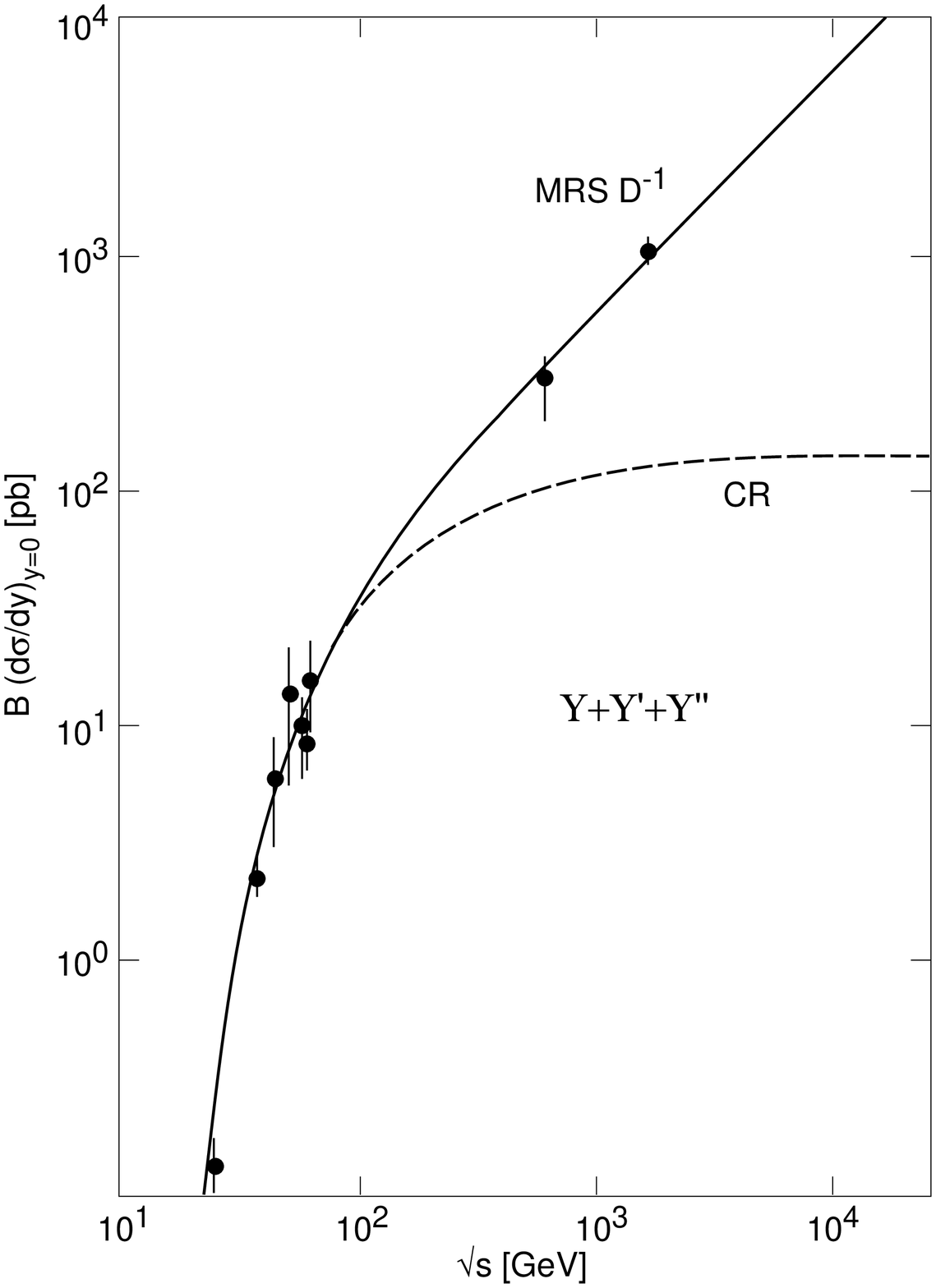,height=0.60\textwidth}}
\end{tabular}
\caption[]{
  Left: The $J/\psi$ production cross section 
  $\sigma^{\p N}[J/\psi]$ for $x_F>0$,
  calculated with MRS D-' PDF, compared to data \cite{Schuler}.
  Right: 
  Energy dependence of $\Upsilon$ production
  $\sum_n \BR[\Upsilon(nS) \rightarrow \mu^+
  \mu^-] \d \sigma[\Upsilon(nS)]^{\p N}/\d y$ at $y=0$ 
  compared to data \cite{Ueno,CDFups}; 
  the predictions with MRS D-' and GRV HO PDF essentially coincide. 
  Also shown (CR) is the phenomenological fit of \cite{Craigie};
  from \cite{Our}.}
\label{figroots}
\end{center}
\end{figure}
\begin{figure}
\begin{center}
\begin{tabular}{cc}
\epsfig{figure=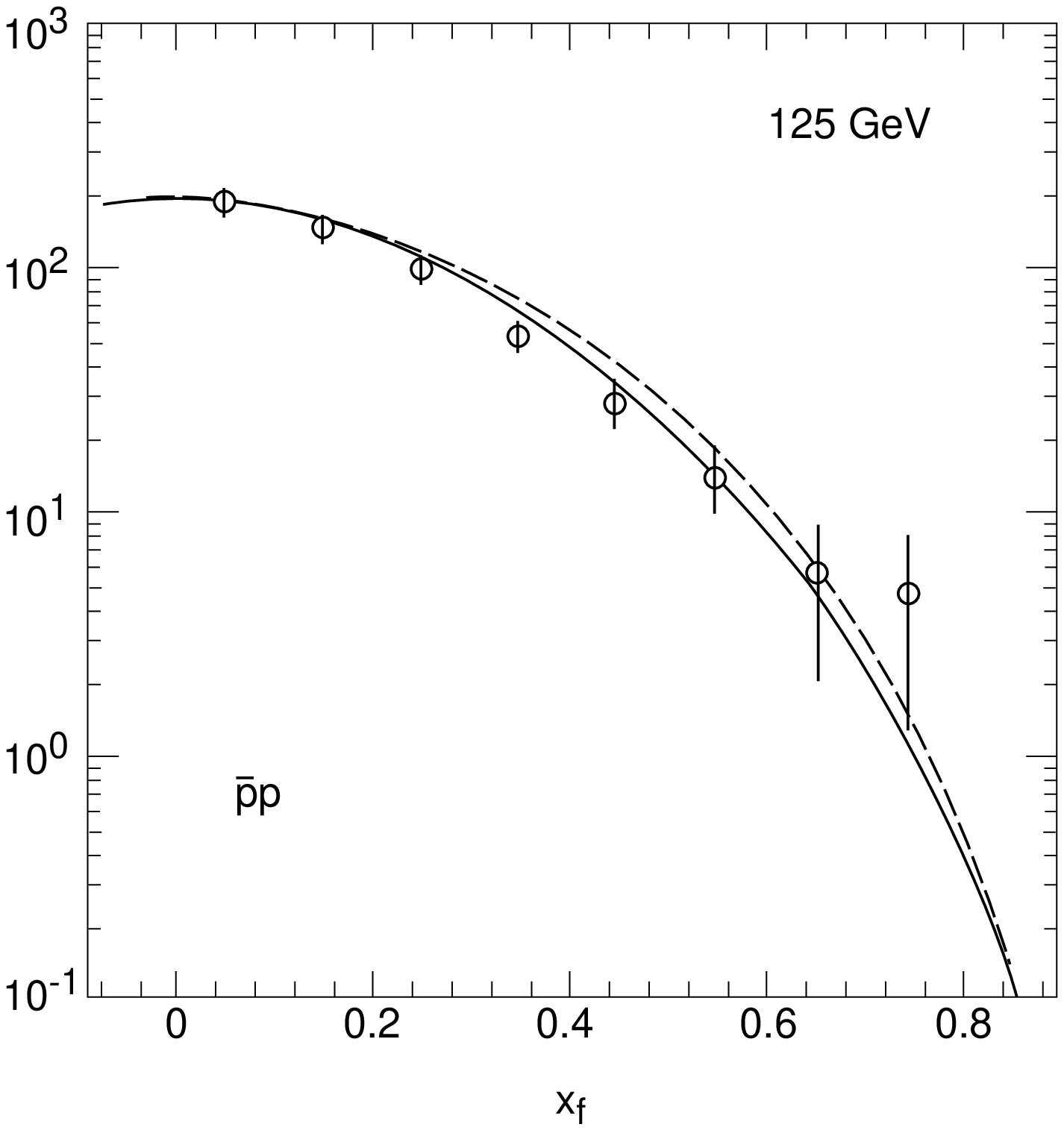,width=0.40\textwidth} &
\epsfig{figure=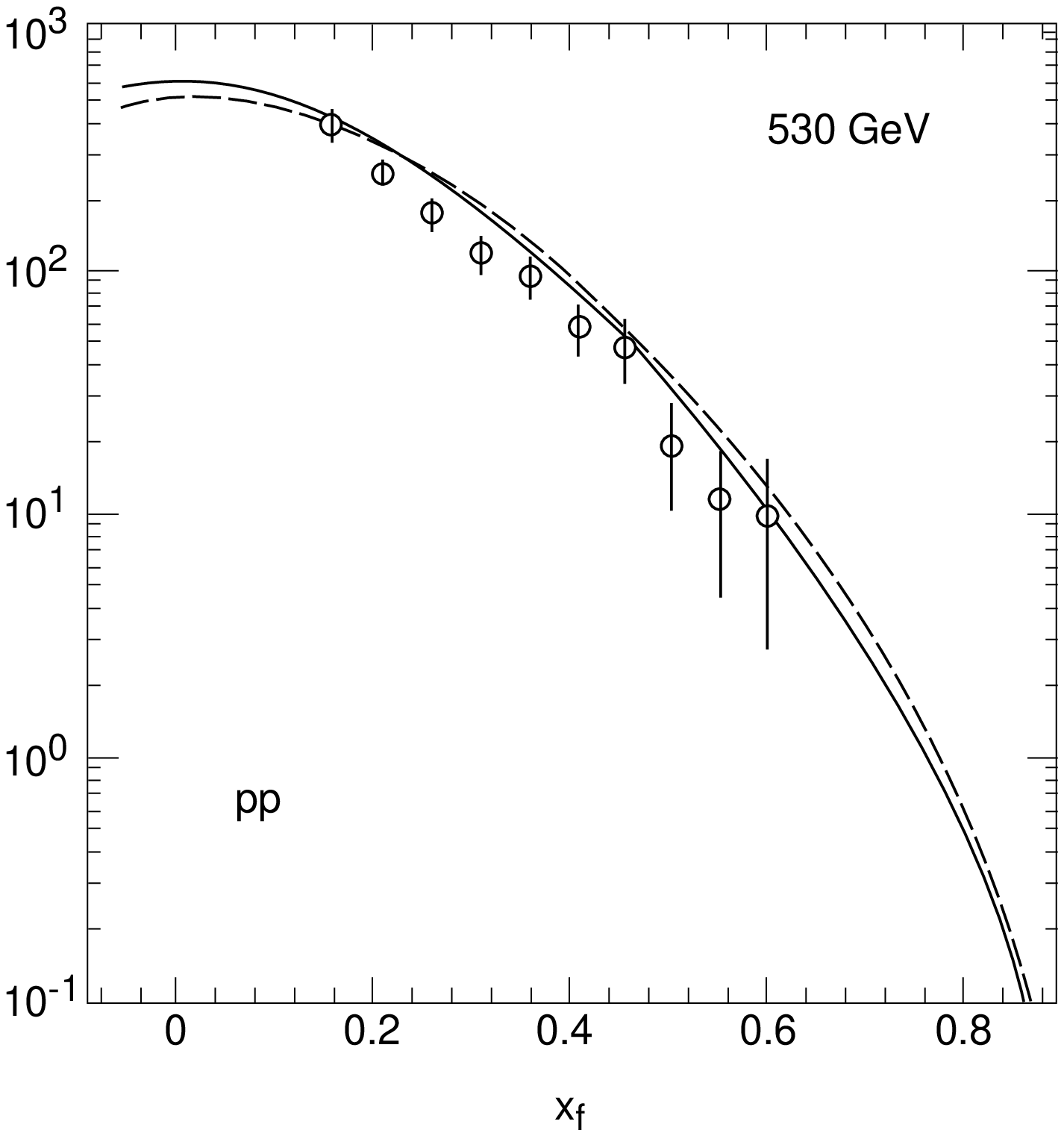,width=0.40\textwidth} \\
\epsfig{figure=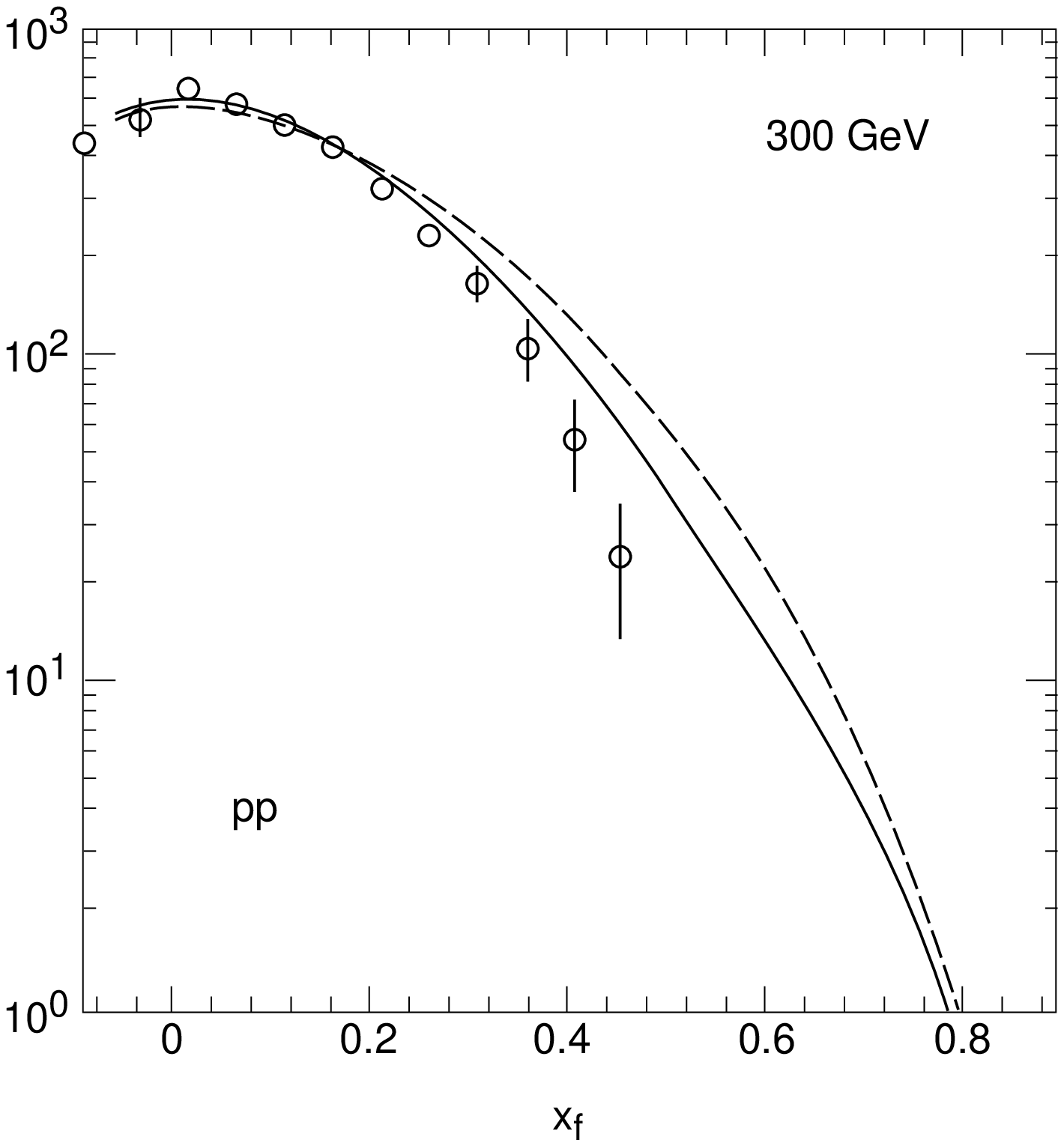,width=0.40\textwidth} &
\epsfig{figure=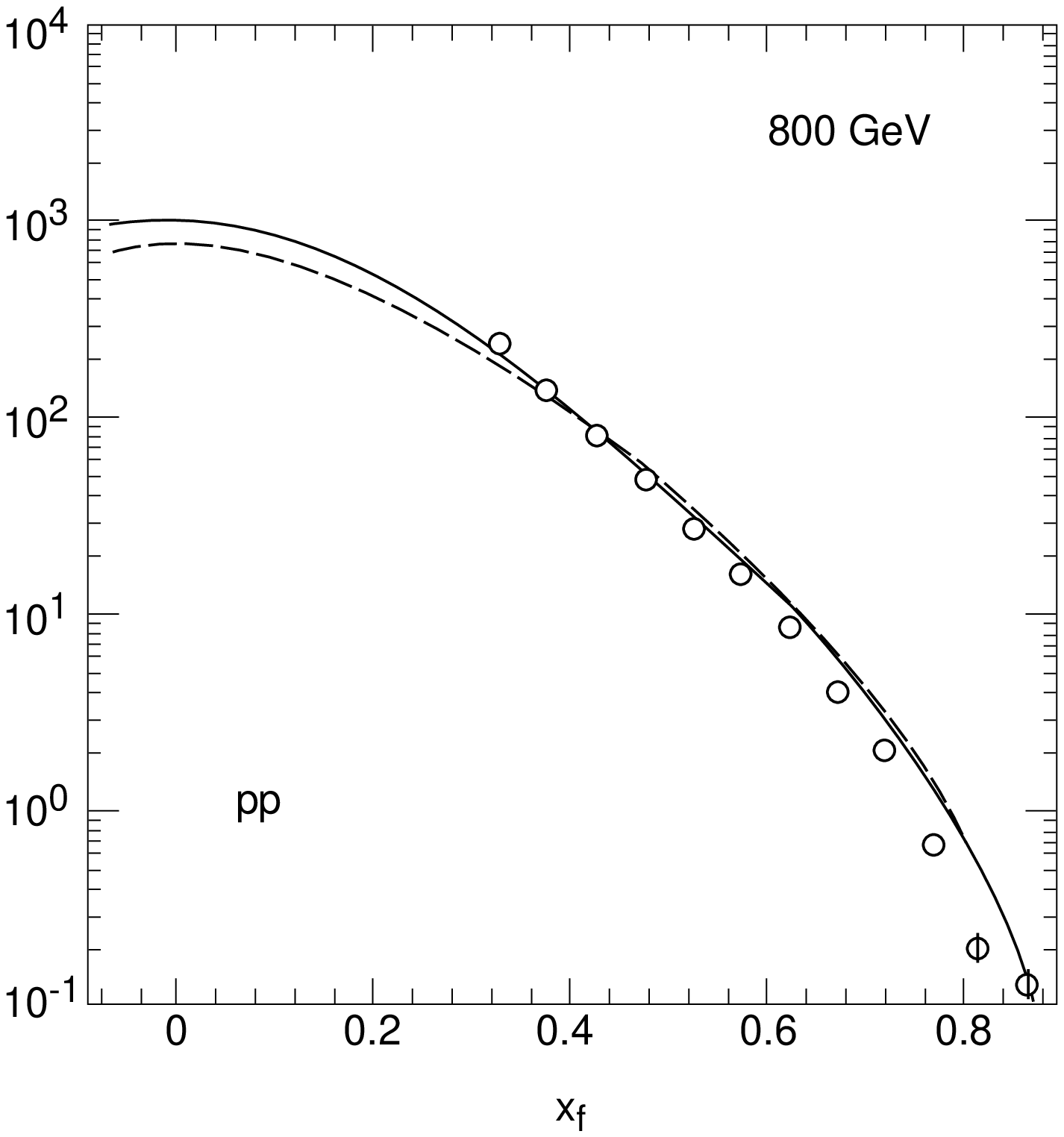,width=0.40\textwidth} 
\end{tabular}
\caption[]{
  The $J/\psi$ longitudinal momentum distributions compared to
  $\pbar N$ and $\p N$ data \cite{x-F}, with $x_F=p_L[J/\psi]/p_{max}[J/\psi]$;
  results obtained with the MRS D-' (GRV) PDF
  are denoted by a solid (dashed) line; from \cite{Our}.}
\label{figxF}
\end{center}
\end{figure}
The short-distance part of eq.\ (\ref{master}) is the perturbative
subthreshold cross section expanded in powers of $\alpha_s(\mu)$ where $\mu
\propto m_Q$.  Specifying to charm, the cross section is
\begin{equation}
  \tilde{\sigma}\left[ \ccbar \right] = 
  \int_{2m_c}^{2m_D}\, \d M_{\ccbar}\, 
   \frac{\d \sigma \left[ \ccbar \right] }{\d M_{\ccbar}}
\ 
\label{subsigma}
\end{equation}
where $\sigma[\ccbar]$ is the {\em spin- and colour-averaged}
open heavy-quark pair production cross section. 
The CEM is hence based on two ingredients. First, the quarkonium dynamics
are assumed to be identical to those of low mass open $Q\bar{Q}$ 
pairs. All perturbative QCD corrections apply to the short-distance
cross section and hence are identical for open and bound heavy-quark
production. Second, although the $\QQbar$ pair is produced at short
distances in different states (distinguished by colour, angular momentum,
relative momentum) and their relative production rates may (and will) be
different for different high-energy collisions, it is only the average 
over many long-distance matrix elements, combined in the long-distance
factor $F[nJ^{PC}]$, that determines the probability to form a specific 
bound state. Necessarily, the factor $F$ needs to be universal, 
i.e.\ process- and kinematics-independent. 

To illustrate the success of the CEM, in Fig.~\ref{figroots} we compare
the prediction for the total $J/\psi$ and $\Upsilon$ production rates 
with data. Note that the
model uniquely predicts the shape of the energy dependence while 
the absolute normalization at low energies fixes the 
non-perturbative factor $F$. Fig.~\ref{figxF} shows the
prediction of fixed-target $x_F$ distributions. There is remarkable agreement 
over a wide energy range, from low-energy $\p\pbar$ collisions where
valence $q \overline q$ fusion dominates up to high-energy 
$\p\p$ collisions dominated by gluon-gluon fusion.  Note that both the shape
and normalization of the $x_F$ distributions are obtained from the model
once $F$ has been fixed by the energy dependence. 

The long-distance factors determined from the low-energy 
total cross sections in Figs.~\ref{figroots} are 
\begin{eqnarray}
  F_{tot}[J/\psi] & = & 2.5\%
\nonumber\\
  \sum_{n=1}^{3}\, {\mrm Br}[\Upsilon(nS) \rightarrow \mu^+ \mu^-]
  \, F_{tot}[\Upsilon(nS)] & = & 1.6 \times 10^{-3}
\ .
\label{Fnumone}
\end{eqnarray}
Here the subthreshold cross sections were calculated in next-to-leading 
order (NLO) using the MRS D-' parametrization \cite{MRS} of the 
parton-distribution functions (PDF) with renormalization and 
factorization scales $\mu_R$ and $\mu_F$ chosen to be 
$\mu_R = \mu_F = 2\, \mc = 2.4\,$GeV and 
$\mu_R = \mu_F = \mb = 4.75\,$GeV, respectively\footnote{To calculate the NLO
subthreshold cross section, we use the program of Mangano, Nason, and Ridolfi
\cite{MNR}, restricting the $Q \overline Q$ mass range.}.
The results in eq.\ (\ref{Fnumone}) are rather insensitive to variations 
of the parameters in the open heavy-quark cross section, if they are
tuned to the open heavy-flavour total cross section data.
For instance, the GRV HO parametrization \cite{GRV} with 
$\mu_R = \mu_F = \mc = 1.3\,$GeV leads to very similar results: 
the smaller subthreshold region is basically compensated by the 
larger two-loop $\alpha_s$ value, $0.298$ for GRV HO
($\Lambda_4 = 0.2\,$GeV) compared to $0.243$ for MRS D-' 
($\Lambda_4 = 0.23\,$GeV). Note, however, that the long-distance factors 
will be considerably larger if the open heavy-quark cross section is 
calculated to leading order (LO) only, a factor $2.2$ larger for $J/\psi$ 
and a factor $1.9$ larger for $\Upsilon$.

\begin{figure}
\begin{center}
\vspace*{-4.0cm}
\epsfig{figure=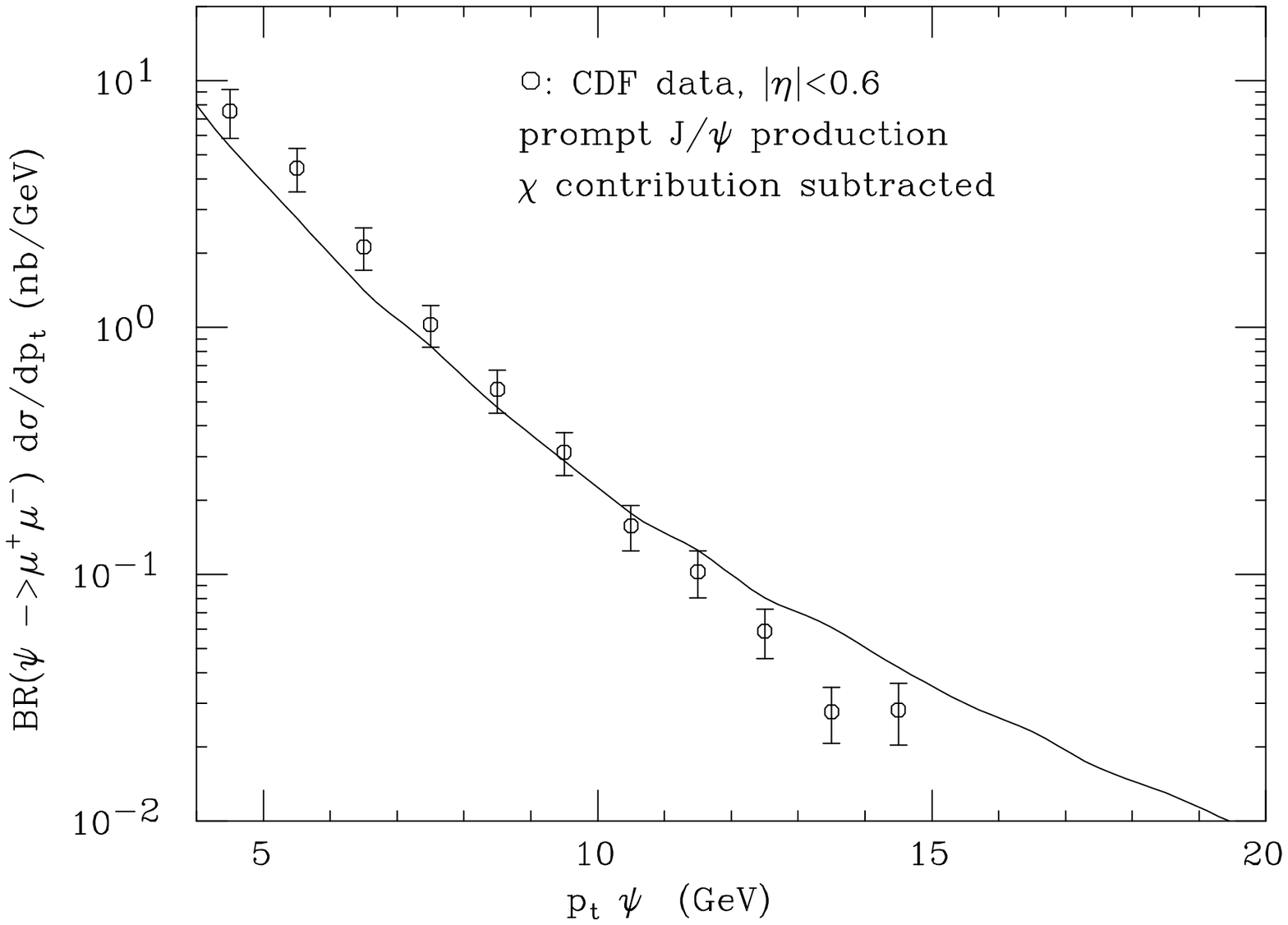,width=0.65\textwidth}%
\\
\vspace*{-6.0cm}
\epsfig{figure=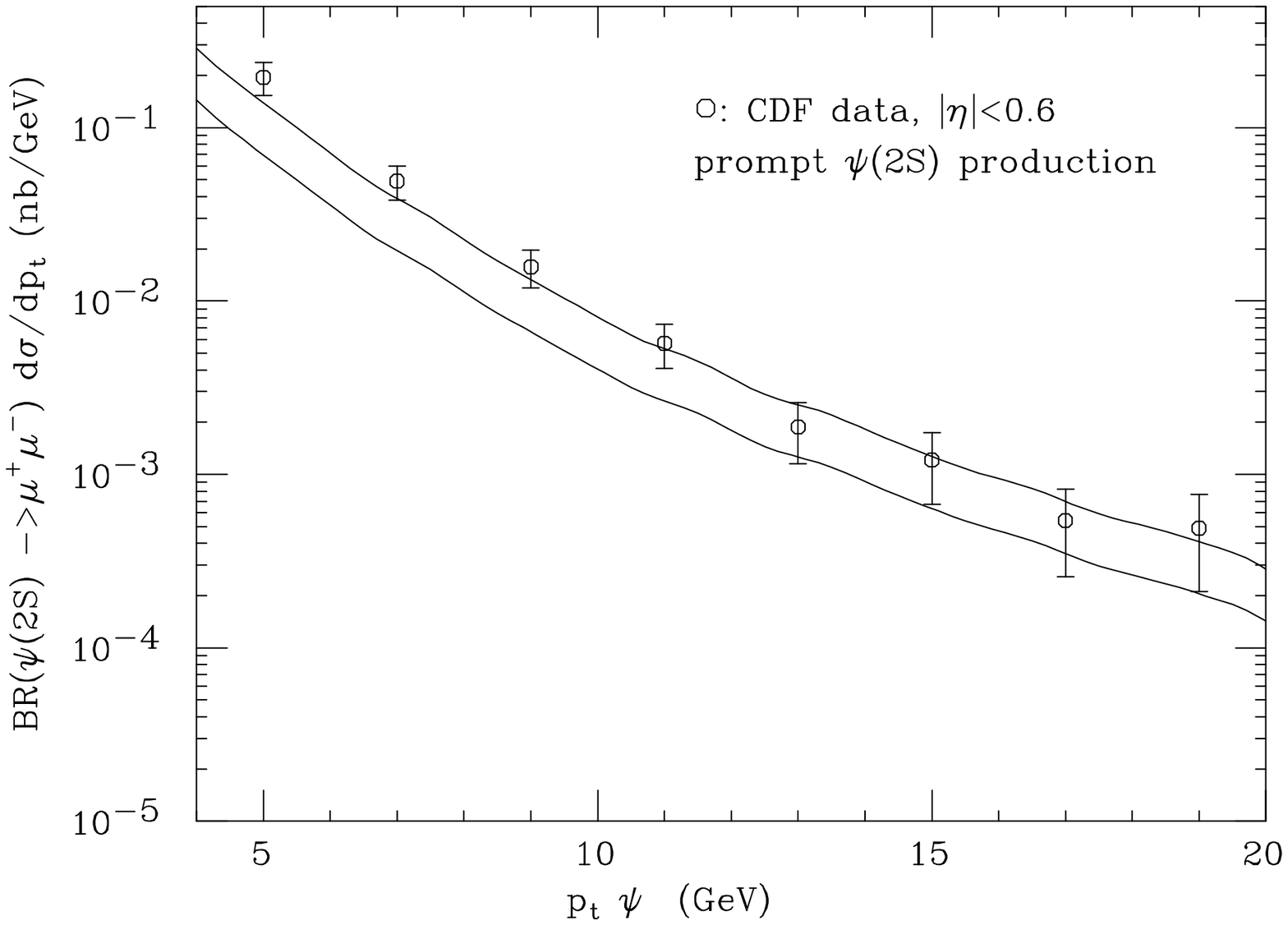,width=0.65\textwidth}%
\\
\vspace*{-6.0cm}
\epsfig{figure=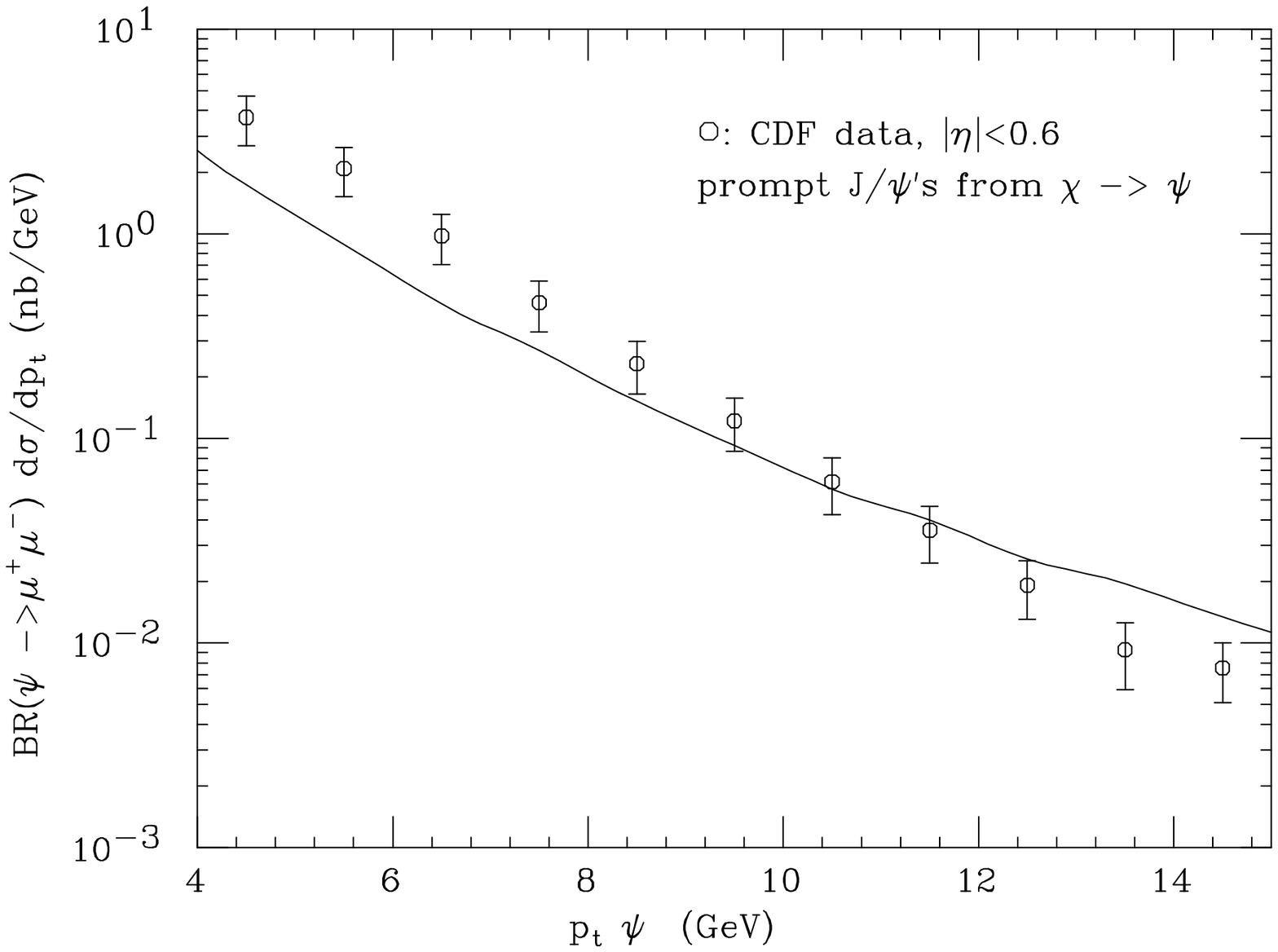,width=0.65\textwidth}%
\\
\vspace*{-3.0cm}
\caption[]{
  Transverse momentum distributions of charmonia compared to CDF 
  data \cite{CDFpsi}.
  The upper curve for $\psi(2S)$ contains an extra $K$-factor of $2$.}
\label{figthree}
\end{center}
\end{figure}
\begin{figure}
\begin{center}
\epsfig{figure=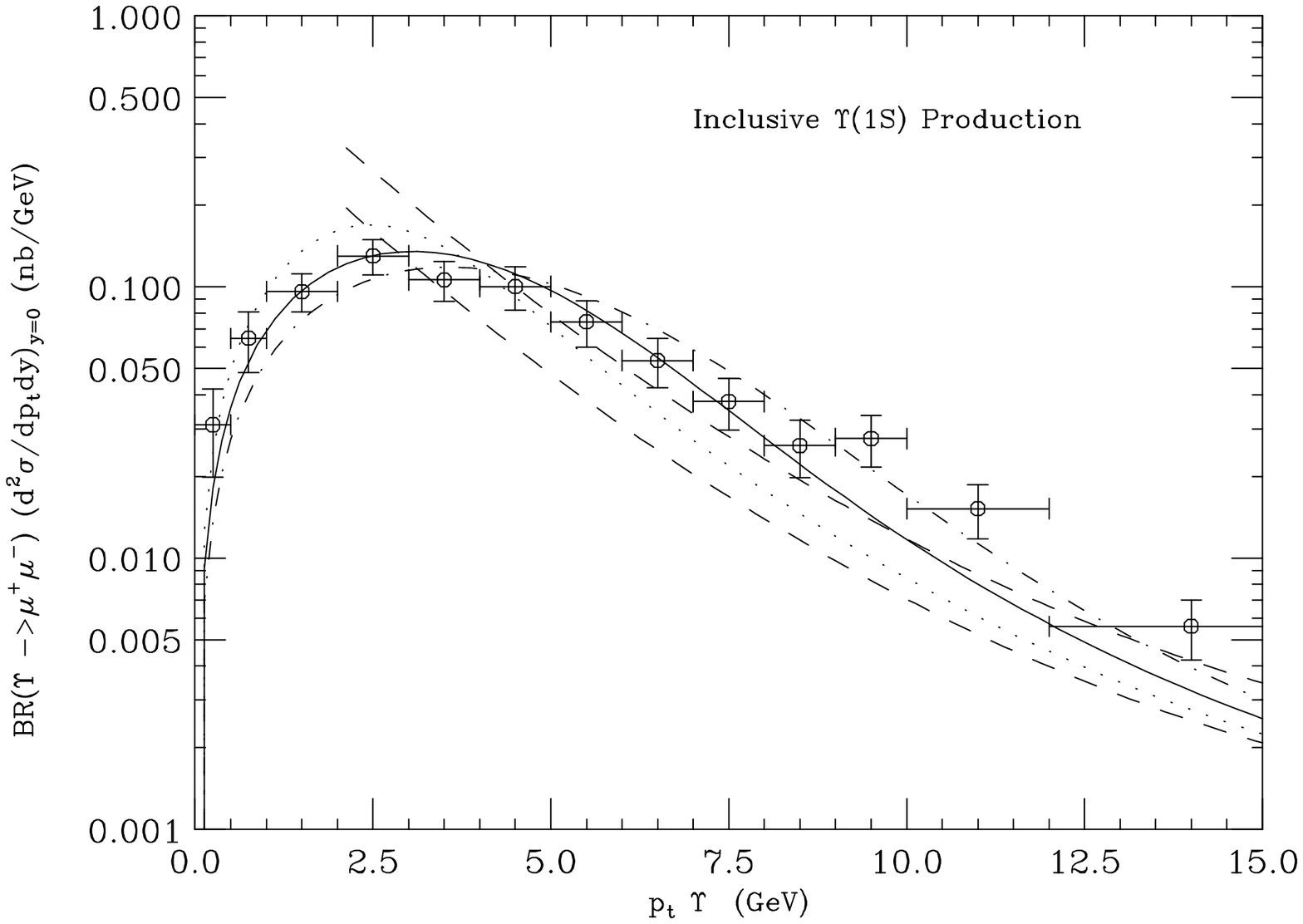,width=0.65\textwidth}%
\\
\epsfig{figure=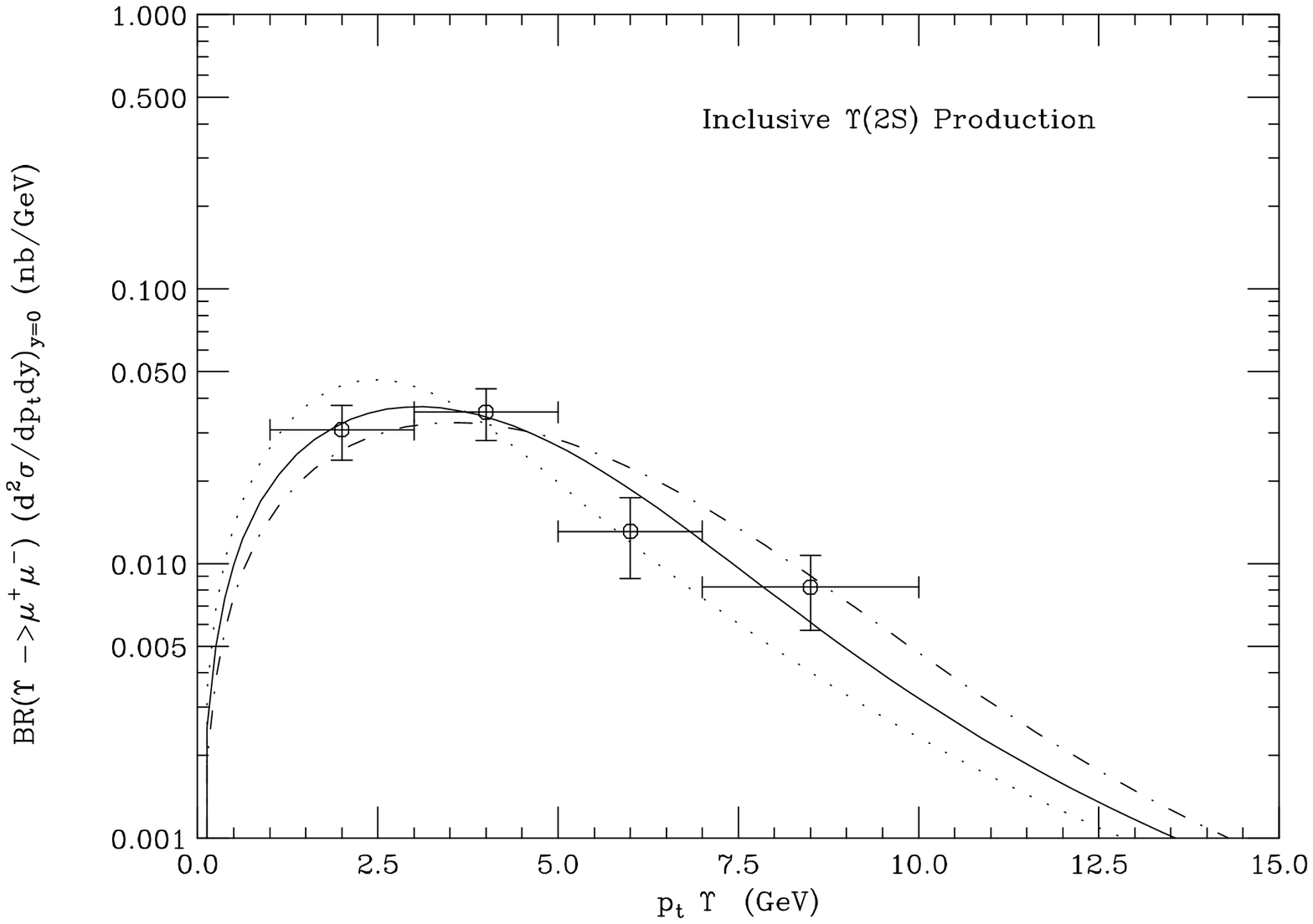,width=0.65\textwidth}%
\\
\epsfig{figure=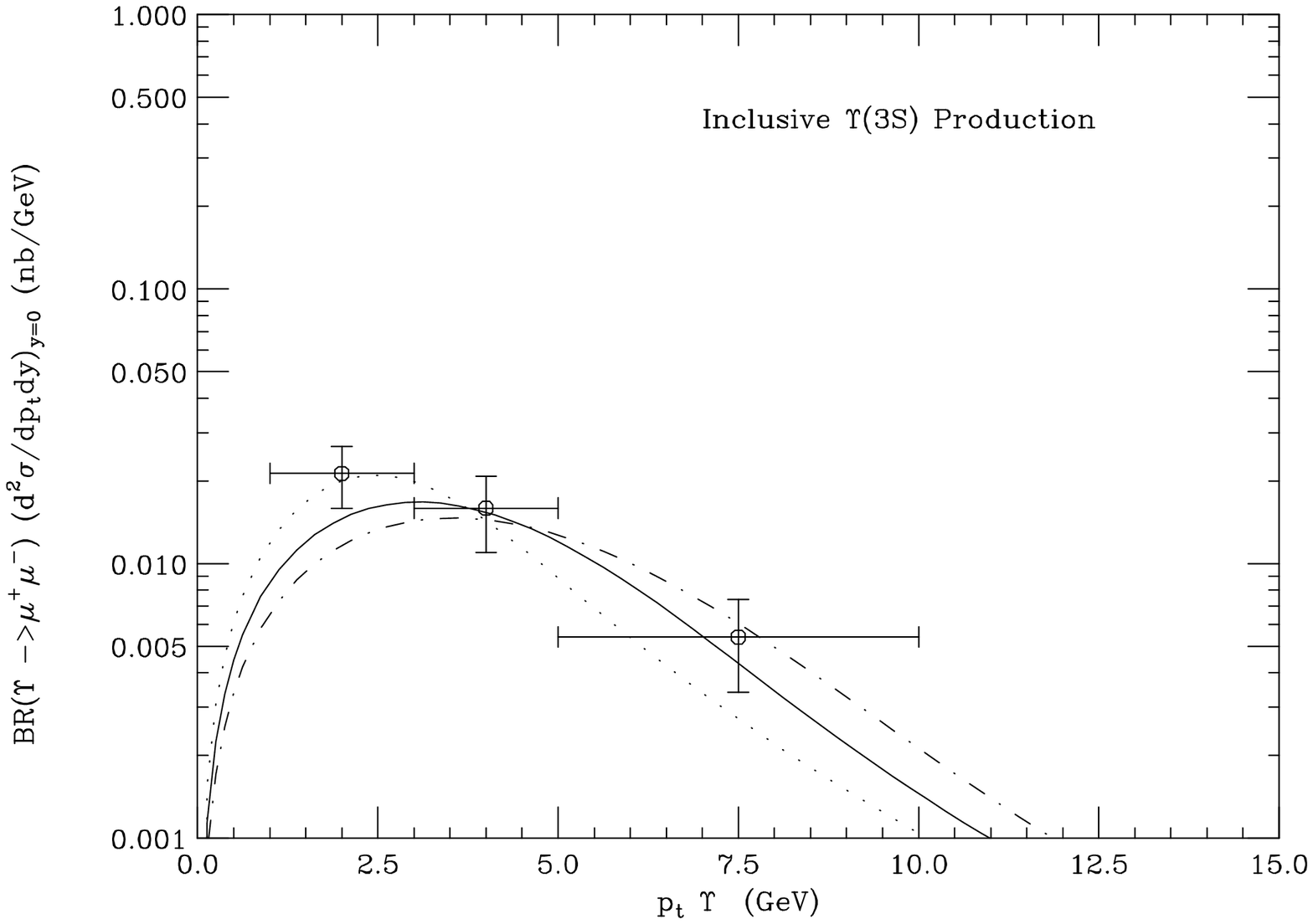,width=0.65\textwidth}%
\\
\caption[]{
  Transverse momentum distributions of bottomonia compared to CDF 
 data \cite{CDFups} for various values of the intrinsic transverse momentum
 with $F$ as in table 1:
 $\langle k_T \rangle = 1.25\,$GeV (dotted), $2.0\,$GeV (solid), 
 $2.5\,$GeV (dot-dashed). The $\Upsilon(1S)$ prediction without smearing
 $\langle k_T \rangle = 0\,$GeV is shown as dashed lines, the upper 
 curve containing an extra $K$-factor of $1.9$.}
\label{figfour}
\end{center}
\end{figure}
In Fig.~\ref{figthree}  we show the transverse
momentum distributions of prompt charmonium 
production (i.e.\ not originating from $B$ decays) at the Tevatron 
energy $\sqrt{s}=1.8\,$TeV.  The bottomonium transverse momentum distributions
are given in Fig.~\ref{figfour}.
The normalizations 
for the various states are given in table~\ref{tableone},
\begin{table}
\renewcommand{\arraystretch}{1.8}
\begin{center}
\begin{tabular}{|c|c|c|c|c|c|}
\hline
   $J/\psi$ & $\psi(2S)$ & $\Upsilon(1S)$ & $\Upsilon(2S)$
 & $\Upsilon(3S)$ & $\sum_{J} {\mrm Br}[\chi_{\c J} \rightarrow J/\psi\, X]\,
  F[\chi_{\c J}]$
\\ \hline
 $2.5$ & $0.35$ & $4.6$ & $2.4$ & $0.78$ & $1.0$
\\ \hline
\end{tabular}
\caption[]{Long-distance factors $F_{\mrm tot}$ from eq.\ (\ref{master}) 
expressed in percent
for $1^{--}$ states and the sum of the inclusive $\chi_{\c J}$
production rates (i.e.\ including cascade decays). 
The factors which multiply the NLO subthreshold cross sections do not include 
the branching ratio into lepton pairs.
For LO calculations the above numbers should be multiplied by $2.2$
for charmonia and $1.9$ for bottomonia.}
\label{tableone}
\end{center}
\end{table}
obtained from eq.\ (\ref{Fnumone}) using the measured cross section ratios 
\cite{Our} and the branching ratio to muon pairs \cite{PDG}.
Good agreement with data is found, typically better than $50$\%. Note that 
the CEM prediction shown in figs.~\ref{figthree} and~\ref{figfour} 
is based on the subthreshold cross section calculated to $O(\alpha_s^3)$, 
which is NLO for the $p_T$ integrated cross section but LO only for 
the $p_T$ distribution\footnote{In the calculation of the $p_T$ distribution,
we used $\mu_R^2 = \mu_F^2 = n^2 [m_Q^2 + (p_{T, Q}^2 + p_{T, 
\overline Q}^2)/2]$ with $n=2$ for charm and 1 for bottom consistent with the
scale advocated for open $Q \overline Q$ production \cite{MNR}.}. 
In the absence of the NLO 
corrections to the $p_T$ spectrum one might apply a $K$ factor to account 
for the unknown higher-order corrections or simply use $F$-values extracted 
in LO (two such examples are given in Figs.~\ref{figthree} 
and~\ref{figfour}). In either case one expects larger cross sections 
so that our estimates are rather conservative.

The CEM prediction for $\psi(2S)$ is about a factor 
of two lower than the data. This is a simple consequence of the fact
that at high $p_T$ the $\psi(2S)$-to-$J/\psi$ ratio measured at the Tevatron
is about twice as large as that observed at fixed-target 
energies for the $p_T$-integrated cross section. Multiplying the CEM 
prediction by a factor of two produces very good agreement. More precise 
data will show whether this is a systematic effect that would require 
refinements of the CEM.

The CEM prediction for the $p_T$ distributions based on fixed-order 
perturbation theory cannot be trusted for $p_T \lessim m_Q$. 
A correct treatment of the low-$p_T$ 
region requires soft-gluon resummation and the inclusion of intrinsic 
transverse momenta, analogous to the Drell--Yan case. The effect of
soft-gluon resummation can be mimicked through an effective, larger 
value of the average intrinsic transverse momentum $\langle k_T \rangle$. 
Fig.\ \ref{figfour} shows that inclusion of $\langle k_T \rangle$ smearing
results in good agreement with data down to very low $p_T$.

The information of table~\ref{tableone} can be used to estimate 
the total bound-state probability. In the case of charmonium this 
requires assumptions about the $\eta_{\c}(nS)$ and $h_{\c}(1P)$ cross
sections. In the case of bottomonium, we also need assumptions
about the ratios of the $\chi_{\b J}(nP)$ to $\Upsilon(nS)$ cross sections. 
Taking the latter to be equal to that measured in the charmonium system and 
assuming that both the $S$- and $P$-wave state cross sections are proportional 
to $2J+1$ and disregarding possible $\bbbar$ $D$-wave states we estimate
\begin{eqnarray}
  \sum_i\, F[i] & \approx ( 8-10)\% & \mrm{charm}
\nonumber\\
                & \approx (17-32)\% & \mrm{bottom}
\ .
\label{Fsum}
\end{eqnarray}
The dominant part of the subthreshold charm cross section 
produces open charm. This fraction is considerably reduced 
in the bottom system: We observe a significant increase 
of the bound state fraction with increasing quark mass. 
The fact that the total charmonium cross section is just $1/(1+8)$ of the
subthreshold cross section
must therefore be considered as fortuitous. This ratio was recently 
advocated as universal for colour-singlet production 
\cite{AEGH96} giving the fraction of both diffractive events in 
deep-inelastic $\e\p$ scatterings and bound states in heavy-quark 
production.  Our analysis shows that bound state production does not obey this
rule.  Moreover there is no reason to expect this ratio to hold.
In fact, considering the complete  system rather than 
restricting to the $Q \overline Q$ pair suggests that the colour-singlet 
fraction is $1:1$ rather than $1:9$ \cite{remark}.

Finally we discuss photoproduction of charmonium. We first note that the 
inelastic $J/\psi$ cross section is defined only within cuts. These cuts are 
necessary since the quasi-elastic process, $\gamma \p \rightarrow J/\psi + 
\p$, and the forward-elastic reaction, $\gamma \p \rightarrow J/\psi + X$, 
where the $J/\psi$ is isolated in rapidity, cannot be described in 
perturbative QCD using a (single) gluon distribution in the nucleon. In 
order to stay away from diffractive production one typically restricts 
$z$ ($E_{\psi}/E_\gamma$ in the nucleon rest frame) to values less than
$0.9$ and applies additional cuts as, e.g., on $p_T$ or on the number of 
tracks.  Since the cross section rises quickly towards $z=1$, $F[J/\psi]$ is 
not well defined here\footnote{A factor of two larger value of $F[J/\psi]$ is
found in \cite{AEGH96}. In the photoproduction case, apparently the 
diffractive processes are not excluded. In the case of hadroproduction, 
the CEM prediction does not seem to contain the $x_F > 0$ constraint
imposed on the data.}. Comparing the cross sections for open charm and
(inelastic) $J/\psi$ photoproduction we find
\begin{eqnarray}
  F[J/\psi] & \approx (1-2.5)\%   &\,\,\, \mrm{at}\,\mrm{low} \, \sqrt{s}
\nonumber\\
            & \approx (0.5-1.4)\% &\,\,\, \mrm{at}\,\mrm{high}\, \sqrt{s}
\ ,
\label{photoratio}
\end{eqnarray}
numbers in reasonable agreement with the value obtained from hadroproduction. 
While the $J/\psi$ long-distance factor thus appears to be universal, 
photoproduction of $\chi_{\c J}$ seems to be problematic for the CEM:
NA14 \cite{NA14} puts an upper limit of about 8\% for the fraction of 
$J/\psi$'s from $\chi_{\c J}$ decays while hadroproduction 
experiments suggest a value four times greater. So far these
measurements are not very precise, but, if confirmed, $\chi_{cJ}$
photoproduction might indicate limitations on the CEM, requiring 
refinements of its simplest variant.

In summary, we find an impressive agreement between data on quarkonium 
production in hadronic collisions and the CEM: Total cross sections, 
$x_F$ distributions, and $p_T$ spectra are all well described by the
assumption that the non-perturbative bound-state formation is governed 
by an average, universal, long-distance factor. Since only a single 
non-perturbative ingredient is required for any given bound state, 
the CEM has great predictive power. Two possible deviations from this 
simple scenario have been pointed out: a difference in the 
$\psi(2S)/J/\psi$ ratio in hadronic collisions and 
a different $\chi_{\c J}$ production fraction in hadro- and photoproduction. 
More data, also for bottomonium, are eagerly awaited.

G.S. thanks the SLAC theory group for hospitality during the completion of this
work.  We also thank H. Satz for many fruitful discussions.




\begin{thebibliography}{99}
%
\bibitem{BFY96}
  For a recent review see e.g.\ 
  E.\ Braaten, S.\ Fleming, and T.C.\ Yuan, OHSTPY-HEP-T-96001 
  \hep{9602374}.
\bibitem{BBL95}
  G.T.\ Bodwin, E.\ Braaten and G.P.\ Lepage, \PRD{51} (1995) 1125.
\bibitem{BSK96}
  G.T.\ Bodwin, D.K.\ Sinclair and S.\ Kim, 
  ANL-HEP-PR-96-28, May 1996 {\it hep-lat/9605023}.
\bibitem{CL96}
  P.\ Cho and A.K.\ Leibovich, \PRD{53} (1996) 150; 
  CALT-68-2026 \hep{9511315}.
\bibitem{BR96}
  M.\ Beneke and I.Z.\ Rothstein, SLAC-PUB-7129 \hep{9603400}.
\bibitem{CK96}
  M.\ Cacciari and M.\ Kr\"amer, DESY-96-005 \hep{9601276};\hfill\\
  J.\ Amundson et al., Univ.\ of Texas preprint UTTG-10-95, 
  \hep{9601298};\hfill\\
  P.\ Ko et al., Univ. of Seoul preprint \hep{9602223}.
\bibitem{Vant}
  M. V\"{a}nttinen, P. Hoyer, S.J. Brodsky, W.K. Tang, \PRD{51} (1995)
  3332;\hfill\\  
  W.K. Tang and M. V\"{a}nttinen, \PRD{53} (1996) 4851;\hfill\\
  W.K. Tang and M. V\"{a}nttinen, NORDITA-96-18-P \hep{9603266}, \PRD{} 
  in press.
\bibitem{Our}
  R.\ Gavai, D.\ Kharzeev, H.\ Satz, G.A.\ Schuler, K.\ Sridhar and
  R.\ Vogt, Int.\ J.\ Mod.\ Phys.\ {\bf A10} (1995) 3043;\hfill\\
  G.A.\ Schuler, CERN preprint, CERN-TH-95-75, April 1995, \hep{9504242}.
\bibitem{Schuler}
  G.A.\ Schuler, ``Quarkonium Production and
  Decays", CERN Preprint CERN-TH.7170/94, February 1994, \hep{9403387}.
\bibitem{Ueno}
  K. Ueno et al., \PRL{42} (1979) 486;\hfill\\
  T. Yoshida et al., \PRD{39} (1989) 3516;\hfill\\
  G. Moreno et al., \PRD{43} (1991) 2815;\hfill\\
  K. Eggert and A. Morsch (UA1), private communication.
\bibitem{CDFups}
  CDF collab., F.\ Abe et al., \PRL{75} (1995) 4358;\hfill\\
  V. Papadimitriou (CDF), 
  Fermilab-Conf-94-221-E, August 1994; 
  Fermilab-Conf-95-227-E, July 1995. 
\bibitem{Craigie}
  N. Craigie, Phys.\ Rep.\ {\bf 47} (1978) 1.
\bibitem{x-F}
  C. Akerlof et al., \PRD{48} (1993) 5067;\hfill\\
  L. Antoniazzi et al., \PRD{46} (1992) 4828;\hfill\\
  M.S. Kowitt et al., \PRL{72} (1994) 1318;\hfill\\
  C. Biino et al., \PRL{58} (1987) 2523;\hfill\\
  V. Abramov et al., ``Properties of $J/\psi$ production in $\pi^--Be$ and
  $p-Be$ Collisions at 530 GeV/c", Fermilab Preprint
  FERMILAB-PUB-91/62-E.
\bibitem{MRS}
  A.D. Martin, R.G. Roberts and W.J. Stirling, \PLB{306} (1993) 145.
\bibitem{GRV}
  M. Gl\"uck, E. Reya and A. Vogt, \ZPC{53} (1993) 127.
\bibitem{MNR}
  M.L.\ Mangano, P. Nason and G. Ridolfi, Nucl. Phys. {\bf B405} (1993) 507.
\bibitem{CDFpsi}
  CDF collab., F.\ Abe et al., \PRL{71} (1993) 2537;\hfill\\
  V.\ Papadimitriou (CDF), Fermilab-Conf-94/136-E (1994); 
  Fermi\-lab\--Conf\--95/\-226-E (1995); 
  Fermi\-lab\--Conf\--96/\-135-E (1996);
  K.\ Ohl (CDF), CDF\-/Pub\-/Bottom\-/Public/\-3676 (1996).
\bibitem{PDG}
  {\it Review of Particle Properties}, L.\ Montanet et al., 
 \PRD{50} (1994) 1173.
\bibitem{AEGH96}
 J.F.\ Amundson, O.J.P.\ Eboli, E.M.\ Gregores, F.\ Halzen,
 Univ.\ of Madison preprint MADPH-96-942, May 1996, \hep{9605295};
 \PLB{372} (1996) 127.
\bibitem{remark}
 T.\ Sj\"ostrand, private communication;\hfill\\
 G.A.\ Schuler, talk presented at the Quarkonium Physics Workshop,
 Univ.\ of Illinois at Chicago, June 1996.
\bibitem{NA14}
 R.\ Barate et al.\ (NA14), \ZPC{33} (1987) 505;\hfill
 P.\ Roudeau (NA14), Nucl.\ Phys.\ Proc.\ Suppl.\ {\bf 7B} (1989) 273.
\end{thebibliography}
\end{document}